# CHAOTIC TEMPERATURES VS COEFFICIENTS OF THERMODYNAMIC ACTIVITY: THE ADVANTAGE OF THE METHOD OF CHEMICAL DYNAMICS

B. Zilbergleyt
Bank One, IT Department, Chicago
E-mail: LIVENT1@MSN.COM

ABSTRACT.
The article compares traditional coefficients of thermodynamic activity as a parameter related to individual chemical species to newly introduced reduced chaotic temperatures as system characteristics, both regarding their usage in thermodynamic simulation of open chemical systems. Logical and mathematical backgrounds of both approaches are discussed. It is shown that usage of reduced chaotic temperatures and the Method of Chemical Dynamics to calculate chemical and phase composition in open chemical systems is much less costly, easier to perform and potentially leads to better precision.

PROBABILISTIC MODEL OF COMPLEX CHEMICAL EQUILIBRIUM.
It is well known how easily the Guldberg-Waage's equation can be derived from the probabilistic considerations. Being based on the particle collisions, typical for the reactions in gases, it states that chance of the reaction to happen is proportional to *joint probability* P(A) of reactant particles to occur simultaneously at the same point of the reaction space. This probability merely equals to product of concentrations of the colliding particles. If multiplied by the rate constant it defines the reaction rate. When the rates of the forward and the reverse reactions are equal then the state of equilibrium exists. Less known is the fact that the above derivation is valid only in the case when no one of reactants can be consumed in any other collision with different outcome than A. Chemical system with only one type of collision represents isolated by the reactants (more traditionally referred to as closed) is a simple chemical system: one type of collision – one outcome.

If more than one type of interactions occur in the system, it becomes complex, and in this case we are interested in probability of collision A given that another collision B (for simplicity, only B besides A) has happened. This kind of a probability of A is referred to as *conditional probability* P(A/B). Because collision B has different than A outcome, both collisions compose a set of complete and mutually exclusive events, and the value of probability P(A/B) is defined by the Bayes' theorem [1]. It is quite enough for the goal of this paper to state that *in general case probability P(A) does not equal to probability P(A/B)*. If only one of reactants participates in collision B, the ratio

$$\gamma = P(A/B)/P(A) \qquad (1)$$

introduces activity coefficient of that reactant. It is quite obvious that $\gamma < = 1$. Now the name of well known Law of *Active* Masses becomes more clear - the value of $\gamma$ defines the fraction of that reactant to participate in collision A. Activity coefficients go into the Guldberg-Waage's equation and then to the equilibrium constant with a product $\gamma \cdot P(A)$.

Thus one can define A and B subsystems of the system under investigation, each of them allow only one type of chemical reaction. In a particular case, if collision B is impossible, subsystem A turns to a closed system. Now outcome of the collision A is independent upon B, hence the probabilities are equal and $\gamma = 1$. This situation defines ideal chemical system.

This model, originally discussed on a different context in [2], is the most abstract and correct way as well to introduce the coefficients $\gamma$ currently known as coefficients of thermodynamic activity. Though the model can be barely used for numeric calculations, it prompts us to conclude that the value of activity coefficient depends not only on the collision A itself but also – and sometimes essentially - upon other possible reactions with this reactant. Using other words, coefficient of thermodynamic activity is to a certain degree an external to A characteristics of the relevant reaction and depends upon composition of the whole system.

With respect to coefficients of thermodynamic activity, the constant equation for j-reaction, following from the probabilistic approach is

$$K_j = \Pi(a_{kj}^{\nu kj}),\qquad(2)$$

or $K_j = \Pi(x_{kj}^{\nu kj}) + \Pi(\gamma_{kj}^{\nu kj})$, where component activities obviously are $a_{kj} = \gamma_{kj} x_{kj}$. Form of the constant equation is unchanged - mole fractions of the reaction participants are replaced by their activities. Nowadays this equation is the basic tool to calculate chemical and phase composition of open chemical systems in equilibrium, at the same time it is supporting the illusion of componental isolation due to holding the same form of constant equation. Knowledge of thermodynamic activity coefficients is a must for thermodynamic simulation of open systems in the current paradigm of chemical thermodynamics.

DYNAMIC MODEL OF COMPLEX CHEMICAL EQUILIBRIUM.
This model deals with equilibrium between an open system and its environment (and has nothing to do with traditional description of dynamic equilibrium on molecular level to describe equilibrium between the forward and the reverse reactions). The initial idea was greatly inspired by d'Alembert's principle [3]. In classical mechanics equilibrium of a mechanical object is always based on equilibrium of the system of mechanical forces, and d'Alembert's principle says "any state of a moving system may be treated as equilibrium state if active forces, acting against the system, are complimented by forces of inertia". Simply speaking, this method turns a non-balanced, open system of forces into complete system. Thus one can handle any non-equilibrium state of mechanical system using equilibrium methods to handle dynamic tasks like static. In application to objects of chemical thermodynamics, similar principle may be worded as follows: any open chemical system (or subsystem of an equilibrium complex system) interacting with its environment (or with its compliment to the whole system) may be suggested as equilibrium if its state is defined by complete set of thermodynamic forces acting against open system. This set includes internal and also external thermodynamic force – a force acting against the open system from its environment. Using thermodynamic affinity in its original meaning as thermodynamic force, this idea forms a background of recently developed Method of Chemical Dynamics [4]. One of the most amazing and practically important results of the method is that it offers an alternative ideology to treat system non-ideality. Basic equation of the method was received in form of

$$\ln [K_j/ \Pi(\eta_{kj}, \Delta^*_j)] + \tau_j \Delta^*_j \delta^*_j = 0,\qquad(3)$$

where $\eta_{kj}$ - thermodynamic equivalent of chemical transformation, $\Delta^*_j$ – reaction extent at open equilibrium, $\delta^*_j$ - shift of the open equilibrium from "true" equilibrium, $\delta^*_j = 1 - \Delta^*_j$, asterisk relates values to open equilibrium, and $\tau_j$ – reduced chaotic temperature (for detailed explanations see [4]). The last value defines resistance of the open chemical system to shift from true thermodynamic equilibrium, and is the most important parameter of the new theory. It was shown in [4] that equation (3) is identical with (2), and in a simple case of one reaction participant, which is common for the j-subsystem and any other part of the subsystem, and $\nu_{kj}= 1$, a very interesting relationship takes place

$$\delta^*_j = (1/\tau_j)[(-\ln \gamma_{kj}^*)/ \Delta^*_j].\qquad(4)$$

There is one principal difference between coefficient of thermodynamic activity and reduced chaotic temperature - parameter $\tau_j$ is essentially internal to the j-subsystem and doesn't depend upon any other reaction in the entire system. Its value can be easily found and tabulated from the stoichiometric equation for any elemental chemical reaction. For example, Table 1 contains values of $\tau_j$ for reaction A+B=C with initial amounts of species of 1, 1, 0 moles, calculated for $\eta_{kj} = 0.6$ (which corresponds to equilibrium constant K= 5.25) and $\Delta^*_j$ varying within the range (0 …1) by the step of 0.05.

Fig. 1 shows graphs of the reaction state shift in open equilibrium from true equilibrium against shifting force F for several values of $\eta_{kj}$.

Basic input suggestion to develop the model was existence of a linear relationship between the reaction shift and shifting force in the vicinity of true equilibrium thus supposing also permanence of the relevant graph slope and its inverse value, $\tau_j$, within a certain, close to true equilibrium region. One can see from the table that, accounting general precision of thermochemical data, the idea very probably will work well up to $\delta_j = 0.3 - 0.4$, and that was used to make calculations in previous publications. On the other hand, better precision can be achieved using tabulated values of $\tau_j$ in relation with the force or reaction shift up to the values of $\delta_j \sim 0.6 - 0.8$, when the curves turn to horizontal.

This example gives an idea how easy is to calculate the value of $\tau_j$ given stoichiometric equation, equilibrium constant or standard change of Gibbs' energy, and initial composition of the reaction mixture.

Table 1.
Reduced chaotic temperatures for elemental reaction A+B=C, $\eta_{kj} = 0.6$ (K = 5.25).

| $F_j$, kJ/m·cd | $\delta_j$ | $\tau_j$ |
|---|---|---|
| 0.00 | 0.00 | |
| 0.18 | 0.05 | 3.68 |
| 0.38 | 0.10 | 3.81 |
| 0.60 | 0.15 | 3.97 |
| 0.83 | 0.20 | 4.16 |
| 1.10 | 0.25 | 4.39 |
| 1.40 | 0.30 | 4.66 |
| 1.75 | 0.35 | 4.99 |
| 2.15 | 0.40 | 5.39 |
| 2.64 | 0.45 | 5.87 |
| 3.24 | 0.50 | 6.47 |
| 3.98 | 0.55 | 7.23 |
| 4.93 | 0.60 | 8.21 |
| 6.19 | 0.65 | 9.52 |
| 7.92 | 0.70 | 11.32 |
| 10.46 | 0.75 | 13.95 |
| 14.46 | 0.80 | 18.07 |
| 21.54 | 0.85 | 25.34 |
| 36.85 | 0.90 | 40.95 |
| 88.52 | 0.95 | 93.18 |

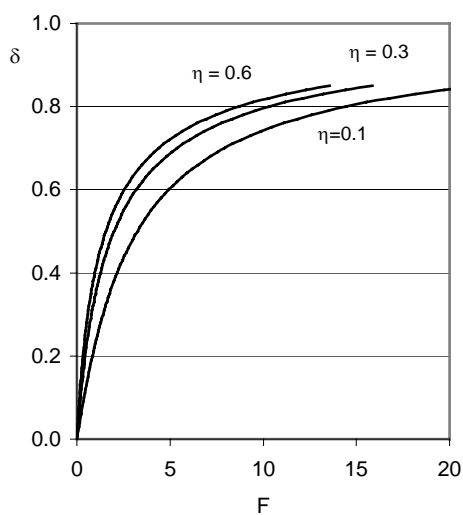

Fig.1. Reaction shift $\delta$ vs. shifting force F, kJ/(m·cd) for elemental reaction A + B = C.

CONCLUSIONS.

In suppressing number of cases coefficients of thermodynamic activity should be determined experimentally – no one even most sophisticated theory can follow their changes in complex chemical systems. Their dependence on the system contents and even upon concentration of the species they are relevant to makes experiments very costly and time consuming. In the contrary, coefficients $\tau_j$ are functions only of the reaction stoichiometry and standard change of Gibbs' free energy. Method of their determination is quite simple and they can be stored in tables for any conceivable stoichiometric equation of elemental chemical reaction. All this provides for a great simplicity and stability on behalf of the Method of Chemical Mechanics.

REFERENCES.